\documentclass[nonatbib]{ragtime}
\title[Modeling an accretion disc stochastical variability]
{Modeling an accretion disc stochastical variability}
\author[T. Pech\'{a}\v{c}ek and V. Karas]
{Tom\'{a}\v{s} Pech\'{a}\v{c}ek
and Vladim\'{\i}r Karas\\
\ins{}Astronomical Institute, 
Academy of Sciences,
Bo\v{c}n\'{\i}~II~1401,
CZ-14131~Prague, Czech~Republic
}
\begin{document}
\begin{abstract}
Hot spots residing on the surface of an accretion disc have been 
considered as a model of short-term variability of active galactic 
nuclei. In this paper we apply the theory of random point processes to
model the observed signal from an ensemble of randomly generated spots.
The influence of general relativistic effects near a black hole is taken
into account and it is shown that typical features of power spectral
density can be reproduced. Connection among spots is also discussed in
terms of Hawkes' process, which produces more power at low frequencies.
We derive a semi-analytical way to approximate the resulting
power-spectral density.
\end{abstract}


\section{Introduction}
Radiation from accreting black holes varies on different timescales 
\citep{gas06}. In X-rays, the observed light-curve, $f\,\equiv\,f(t)$, is a 
complicated noisy curve that can be represented by a broad-band power 
spectrum \citep{2002MNRAS.332..231U}. It has been proposed
\citep{Abramowicz:1991:A&A,1992AIPC..254..251W} that `hot spots' are a
possible contributor to this variability. These spots are supposed to
occur on the surface of an accretion disc following its irradiation by
coronal flares 
\citep{Galeev:1979:APJ,2001MNRAS.328..958M,2004A&A...420....1C}.
A model light-curve can be constructed as a sum of contributions from
many point-like sources that are orbiting above an underlying accretion
disc. The observed signal is modulated by relativistic effects as
photons propagate towards a distant observer.

In order to characterise the light curves we need to introduce some
appropriate estimator of the source variability. In a mathematical
sense, one applies a functional: $f\rightarrow
\mathcal{S}\left[f\right]$, where $\mathcal{S}\left[.\right]$ is a
map from functions defined on $\mathbb{R}$ to functions on
$\mathbb{R}^k$ ($k\geq 0$). The variability estimator can be a single
number (for example, the mean flux or the `rms' characteristic), or
function of one variable (power spectrum density or probability
distribution) or of many variables (poly-spectra, rms--flux relation,
etc). A signal of such a spotted accretion disc should be
intrinsically stochastical. Hence, the variability
estimator $\mathcal{S}\left[f\right]$, derived from a piece of the
light-curve, is a random value, too.

Various schemes have been proposed in which spots are mutually
interconnected in some way \citep{Poutanen:1999:MNRAS,
Zycki:2005:MNRAS}. We want to investigate this type of models within a
common mathematical basis. Other authors have developed different
approaches  to the problem
\citep{1994PASJ...46...97M,1997MNRAS.292..679L}.

\section{Spot models and the accretion disc variability}
\subsection{Model assumptions and variables}
Let us have $K$ samples of the observed light-curves from the same 
source, $f_j$. The law of large numbers ensures that
\begin{equation}
\frac{1}{K}\sum\limits_{j=1}^K\mathcal{S}\left[f_j\right]
\rightarrow {\rm E}\left[\mathcal{S}\left[f\right]\right],
\quad K\rightarrow\infty,
\end{equation}
where ${\rm E}[.]$ is the mean value operator. The average value of the
functional is formally defined
\begin{equation}
{\rm E}\left[\mathcal{S}\left[f\right]\right]
\equiv\sum\limits_{\{{\rm All\, possible\,} f_j(t)\}}
\left({\rm Probability\, of\, }f_j\right)
\times \mathcal{S}\left[f_j\right],
\end{equation}
where the sum goes over all possible light-curves generated by this
model. We will show how to define and parameterise ``the space of all
possible light-curves'' and how to perform the averaging when the
functional is the power spectrum. 

\begin{figure*}
\begin{center}
\includegraphics[width=0.8\textwidth]{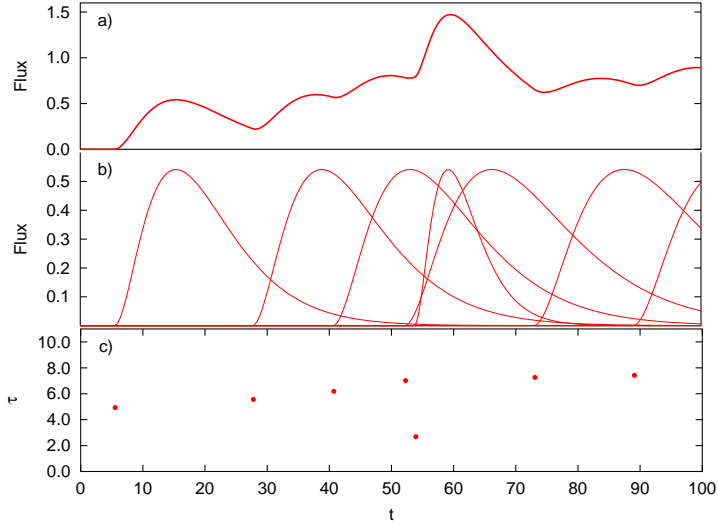}
\caption{The model light-curve (panel a) is obtained as a sum of elementary 
events (panel b). Profile of each individual event is assumed to be 
$I(t,\tau)=\left(t/\tau\right)^2 \exp\left(-t/\tau\right)\theta(t)$,
as described in the text. Their normalization is identical.
The final light-curve is fully determined by the form of the individual 
contributions together with a set of points in $t$--$\tau$ plane (panel c), 
which represent pairs of ignition times and temporal constants
$\tau$ of each event.}
\label{krivex2}
\end{center}
\end{figure*}

The general model is constrained only by the following three assumptions
about the creation and evolution of spots:
\begin{description}
\item (i)~Each spot is described by its time and place of birth 
($t_j$, $r_j$ and $\phi_j$) on the disc surface. 
\item (ii)~Every new occurrence starts instantaneously; afterwards the emissivity 
decays gradually to zero. The total emitted radiation energy is finite. 
\item (iii)~The intrinsic emissivity can be fully
determined by a finite number of parameters which form a vector
$\mbox{\boldmath $\xi$}_j$.
\end{description}
For a simple demonstration of this concept see figure \ref{krivex2}.
The disc itself has a passive role in our present considerations.
We will treat it as a geometrically thin, optically thick layer lying
in the equatorial plane.

\subsection{Random point processes}
The concept of point processes is a generalisation of well-known random
processes which were developed as a description of time-dependent 
random values \citep{Bendat:2000}. Point processes are used as statistical 
description of configurations of some randomly distributed points in space 
$\mathbb{R}^n$.

One way of describing a configuration of points is by their
counting measure, $N(A)$, which for every $A \subset \mathbb{R}^n$
gives a number of points lying in $A$. One defines the intensity measure,
\begin{equation}
M_1(A)={\rm E}\left[N(A)\right].
\end{equation}
Similarly to random processes,
the point process can be characterised by its {\em mean value} and
{\em moments}.
For every $A\subset\mathbb{R}^n$, $M_1(A)$ is the mean number of points lying
in $A$. The second-order moment measure is defined in the same way
on the Cartesian product of spaces $\mathbb{R}^n\times\mathbb{R}^n$:
\begin{equation}
M_2(A\times B)={\rm E}\left[N(A)N(B)\right].
\end{equation}
Let ${x_i}_N$ be one possible configuration of points, i.e.\ the support of
some $N(.)$ For the functions $f(x)$ and $g(x,\, y)$ on $\mathbb{R}^n$
and $\mathbb{R}^{2n}$, respectively, it follows \citep{Campbell:1909,Daley:2003}
\begin{eqnarray}
{\rm E}\left[\ \sum\limits_{\{x_i\}_N}f(x_i)\right]
&=&{\rm E}\left[\ \int\limits_\mathcal{X}f(x)N({\rm d}x)\right]
\;=\;\int\limits_{\mathbb{R}^n}f(x)M_1({\rm d}x)\label{Camb1}
\\{\rm E}\left[\ \sum\limits_{\{x_i\}_N,\,\{y_i\}_N}g(x_i,\,y_i)\right]
&=&{\rm E}\left[\ \int\limits_{\mathbb{R}^{2n}}g(x,\,y)N({\rm d}x)N({\rm d}y)\right]
\nonumber\\
&=&\int\limits_{\mathbb{R}^{2n}}g(x,\,y)M_2({\rm d}x\times{\rm dy}).
\label{stred2}
\end{eqnarray}

The concept of point process can be further generalised in the following
way. We add a {\em mark} $\kappa_i$ from the mark set $\mathcal{K}$ to each
coordinate $x_i$ from $\{x_i\}_N$. Marks carry additional
information. The resulting  point process on the set
$\mathbb{R}^n\times\mathcal{K}$ is called the `marked point process' if for
every $A\subset\mathbb{R}^n$ it fulfills the condition  $N_{\rm
g}(A)\equiv N(A\times\mathcal{K})<\infty$. 

The random measure $N_{\rm g}(A)$ represents the {\em ground process} of
the marked process $N$. When the dynamics of the process is governed
only by the ground process and marks are mutually independent and
identically distributed random values with the distribution functions
$G({\rm d}\kappa)$, then the process intensity and the second order
measure fulfill
\begin{eqnarray}
M_1({\rm d}x\times{\rm d}\kappa)&=&M_{1{\rm g}}({\rm d}x)G({\rm d}\kappa),\label{markpp1}\\
M_2({\rm d}x_1\times{\rm d}\kappa_1\times{\rm d}x_2\times{\rm d}\kappa_2)
&=&M_{{\rm g}2}({\rm d}x_1\times{\rm d}x_2)G({\rm d}\kappa_1)G({\rm d}\kappa_2)\label{markpp2}.
\end{eqnarray}

\subsection{Relationship between point processes and spots}
Let us assume a surface
element orbiting at radius $r$ with constant emissivity $I$ and
orbital frequency $\Omega(r)$. This should represent an infinitesimally
small spot. For the flux measured by an
observer at inclination $\theta_{\rm o}$ we find
\begin{equation}
f(t)=IF(t,\,r,\,\theta_{\rm o}).
\end{equation}
The periodical modulation of the signal is determined by relations
\begin{eqnarray}
F(t(\phi),\,r,\,\theta_{\rm o})=F(\phi,\,r,\theta_{\rm o}),\\
t(\phi)=\frac{\phi}{\Omega(r)}+\delta t(\phi,\,r,\theta_{\rm o}),
\end{eqnarray}
where $F(\phi,\,r,\theta_{\rm o})$ is the transfer function describing the total
amplification of signal emitted from then disc surface element on the coordinates 
$r$ and $\phi$. The function $\delta t(\phi,\,r,\theta_{\rm o})$ is the time delay
of the signal (hereafter we will omit $\theta_{\rm o}$ in the argument of $F$ for 
simplicity).
Now, we consider a process consisting of statistically dependent events,
\begin{equation}
f(t)=\sum\limits_j I(t-{\delta_j},\,\mbox{\boldmath $\xi$}_j)F(t-\delta_{{\rm p}j},\, r_j),
\label{proces}
\end{equation}
where: $I(t,\,\mbox{\boldmath $\xi$})=\theta(t)g(t,\, \mbox{\boldmath
$\xi$})$  is the profile of a single event; $\delta_j=t_j+t_{0j}$ is
time offset; $\delta_{{\rm p}j}=\delta_j+t_{{\rm p}j}$ is the phase 
offset; $\theta(t)$ is the Heaviside 
function; and $g(t,\,\mbox{\boldmath $\xi$})$ is non-negative function of $k+1$
variables $t$ and $\mbox{\boldmath $\xi$}=(\xi^1,\dots,\xi^k)$, which is
on  the interval $\langle0,\,\infty)$ integrable in the variable $t$ for
all values of parameters $\mbox{\boldmath $\xi$}\in\Xi$. The set $\Xi$
is some measurable subset of $\mathbb{R}^k$. For a fixed value of $r$,
$F(t, \, r)$ is a periodical function of $t$, with the angular frequency
$\Omega(r)$. 

Quantities $\mbox{\boldmath $\xi$}_j$, $t_j$, $r_j$,  $t_{{\rm p}j}$ and
$t_{0j}$ are random values. The vector $\mbox{\boldmath $\xi$}_j$
determines  the  duration and shape of each event, $t_j$ is time of
ignition of the $j$--th event, and $t_{0j}$ the corresponding initial
time-offset. Parameter $t_{{\rm p}j}$ determines the initial phase of
the periodical modulation. Processes of this kind and their power 
spectra were mathematically studied by Br\'{e}maud and Massouli\'{e},
\citep{Bremaud:2002:AAP, Bremaud:2005:AAP}.

Power spectral function of a stationary stochastic process $X(t)$ is
\begin{equation}
S(\omega)=\lim\limits_{T\rightarrow\infty}\frac{1}{2T}
{\rm E}\left[\left|\mathcal{F}_T[X(t)](\omega)\right|^2\right],
\label{powersp}
\end{equation}
where $\mathcal{F}_T[\,]$ is the incomplete Fourier transform,
\begin{equation}
\mathcal{F}_T[X(t)]=\int\limits^T_{-T}X(t)e^{-i\omega t}{\rm d}t.
\label{cast_four}
\end{equation}
This can be evaluated by using the complete Fourier transform,
\begin{equation}
\int\limits^T_{-T}X(t)e^{-i\omega t}{\rm d}t
=\int\limits^\infty_{-\infty}X(t)\chi_{{\langle-T,\,T\rangle}}(t) e^{-i\omega t}{\rm d}t=
2\frac{\sin(T\omega)}{\omega}\star\mathcal{F}[X(t)](\omega),
\end{equation}
where $\chi_{A}(x)$ is the characteristic function of set $A$, which equals $1$ for $x\in A$
and $0$ for $x\not\in A$.
Symbol $\star$ denotes the convolution operation.

By applying this transformation on the process (\ref{proces}) we find
\begin{equation}
\mathcal{F}_T[f(t)](\omega)=
\frac{2\sin(T\omega)}{\omega}\star\sum\limits_j
\mathcal{F}[I(t-{\delta_j},\,\mbox{\boldmath $\xi$}_j)\;F(t-\delta_{{\rm p}j},\, r_j)](\omega).
\end{equation}
The Fourier transform of a single event 
$I(t-{\delta_j},\,\mbox{\boldmath $\xi$}_j)F(t-\delta_{{\rm p}j},\, r_j)$
is then
\begin{equation}
\mathcal{F}[I(t-{\delta_j},\,\mbox{\boldmath $\xi$}_j)F(t-\delta_{{\rm p}j},\, r_j)](\omega)
=e^{-i\omega\delta_j}\mathcal{F}[I(t,\,\mbox{\boldmath $\xi$}_j)]\star\mathcal{F}[F(t+t_{{\rm p}j},\, r_j)].
\end{equation}
Function $F(t,\,r)$ is periodical in time, and so it can be expanded:
\begin{equation}
F(t,\, r)=\sum\limits^\infty_{k=-\infty}c_k(r) e^{ik\Omega(r) t},
\end{equation}
where $\Omega(r)$ is the frequency of $F(t,\, r)$. We find
\begin{eqnarray}
\mathcal{F}\left[F(t_{\rm p},\, r)\right](\omega)&=&\sum\limits_{k=-\infty}^\infty
c_k(r) e^{ik\Omega(r) t_{\rm p}}\delta\left(\omega-k\Omega(r)\right),\\
\mathcal{F}\left[I(t,\,\mbox{\boldmath $\xi$})\right] \star \mathcal{F}\left[F(t+t_{\rm p},\, r)\right]
&=&\sum\limits_{k=-\infty}^\infty c_k(r) e^{ik\Omega(r) t_{\rm p}}
\mathcal{F}\left[I(t,\,\mbox{\boldmath $\xi$})\right]\left(\omega-k\Omega(r)\right).
\label{konvoluce}
\end{eqnarray}
The above given formulation of the problem falls perfectly within the mathematical 
framework of point processes.

\subsection{The case of independent decaying spots (Poisson process)}
Knowing the incomplete Fourier transform of $f(t)$ we can now calculate 
its squared absolute value and perform the averaging over all 
realizations of the process. Between $-T$ and $T$ the process is influenced 
by all events ignited during the preceding interval
$\langle-\infty,\, T \rangle$, however (because of fast decay of
every single event), this can be restricted onto
$\langle-(T+C),\, T \rangle$, where $C$ is a sufficiently large
positive constant. Therefore, every realization of the process $f(t)$ on the interval 
$\langle-T,\, T \rangle$ can be described by set of points in 
$(k+4)$--dimensional space $(t_j,\,t_{0j},\,t_{{\rm p}j},\, r_j,\,\mbox{\boldmath $\xi$}_j)$, 
where $t_j \in \langle-(T+C),\, T \rangle$.

Equation (\ref{proces}) represents a very general class of random processes.
However, in all reasonable models of spotted accretion discs the values of
initial time delay and phase are functions of initial position of each spot
($r$ and $\phi$), i.e.\
\begin{equation}
t_0=\delta t(r,\, \phi),\qquad
t_{\rm p}=\frac{\phi}{\Omega(r)}+t_0.
\end{equation}
Fourier transform of the resulting signal can be then simplified,
\begin{eqnarray}
\mathcal{F}[I(t-t_{0j},\,\mbox{\boldmath $\xi$}_j)&&
F(t-t_{0j}+t_{{\rm p}j},\, r_j)](\omega)\nonumber\\
&&=\sum\limits_{k=-\infty}^\infty
c_k(r)e^{ik\phi}\mathcal{F}\left[I(t-\delta t(r,\,\phi),\,
\mbox{\boldmath $\xi$})\right]\left(\omega-k\Omega(r)\right).
\end{eqnarray}
Every realization of this process is completely determined
by set of points $(t_j,\,\phi_j,\, r_j,\,\mbox{\boldmath $\xi$}_j)$
from some subset of $\mathbb{R}^{k+3}$.

For the sum of $K$ complex numbers $z_i$ it follows
\begin{equation}
\left|\sum\limits_{i=1}^K z_i\right|^2=
\left(\sum\limits_{i=1}^K z_i\right)\left(\sum\limits_{i=1}^K z_i\right)^*
=\left(\sum\limits_{i=1}^K z_i\right)\left(\sum\limits_{i=1}^K z_i^*\right)
=\sum\limits_{i=1}^K\sum\limits_{j=1}^K z_i z_j^*.
\label{complexnumbers}
\end{equation}
Defining the function $s(t,\phi,r,\mbox{\boldmath $\xi$};\omega)$ as
\begin{equation}
s(t,\phi,r,\mbox{\boldmath $\xi$};\omega)=
\frac{2\sin(T\omega)}{\omega}\star
\left(e^{-i\omega t}\!\sum\limits_{k=-\infty}^\infty \!c_k(r) e^{ik\phi}
\mathcal{F}\left[I(t-\delta t,\,\mbox{\boldmath $\xi$})\right](\omega-k\Omega(r))\right).
\label{sfunkce1}
\end{equation}
According to (\ref{complexnumbers}) we can write
\begin{eqnarray}
\left|\mathcal{F}_T[f(t)](\omega)\right|^2
&=&\left|\sum\limits_j s(t_j,\,\phi_j,\,r_j, \,\mbox{\boldmath $\xi$}_j;\,\omega)\right|^2\nonumber\\
&=&\sum\limits_j \sum\limits_l s(t_j,\,\phi_j,\,r_j, \,\mbox{\boldmath $\xi$}_j;\,\omega)\,
s^*(t_l,\,\phi_l,\,r_l, \,\mbox{\boldmath $\xi$}_l;\,\omega).
\end{eqnarray}
Due to Campbell's theorem (\ref{stred2}),
\begin{eqnarray}
{\rm E}\left[\left|\mathcal{F}_T[f(t)](\omega)\right|^2\right]=
{\rm E}\left[\sum\limits_j \sum\limits_l s(t_j,\,\phi_j,\,r_j, \,\mbox{\boldmath $\xi$}_j;\,\omega)\,
s^*(t_l,\,\phi_l,\,r_l, \,\mbox{\boldmath $\xi$}_l;\,\omega)\right]\nonumber\\
=\int\limits_{A\times A'} s(t,\,\phi,\,r,\,\mbox{\boldmath $\xi$};\,\omega)\,
s^*(t',\,\phi',\, r',\,\mbox{\boldmath $\xi$}';\,\omega)\,
m_2(t,\,\phi,\,r,\mbox{\boldmath $\xi$},\,t',\,\phi',\,r',\,\mbox{\boldmath $\xi$}')\,{\rm d}A\,{\rm d}A',
\label{Eft2}
\end{eqnarray}
where $m_2$ is density of the second-order moment measure corresponding to the random point process of 
$(t_j,\,\phi_j,\,r_j,\, \mbox{\boldmath $\xi$}_j)$. The set $A$ is a Cartesian product of sets,
\begin{equation}
A=\langle-(T+C),\, T\rangle \times \langle 0,\, 2\pi\rangle\times
\langle r_{\rm min},\, r_{\rm max}  \rangle\times\Xi.
\end{equation}

Now we can perform the limit (\ref{powersp}). It can be shown that the result is independent
on the value of $C$. In order to obtain an explicit formula for the power spectral density
we have to specify the form of $M_2(.)$. In the simplest case we assume events that are mutually 
independent with uniformly distributed ignition times.
The process can be described as a marked point process with a Poissonian process as the ground process.
The intensity and the second-order measure for the ground process are:
\begin{eqnarray}
M_{{\rm g}1}({\rm d}t)&=&n{\rm d}t,\label{PoissLambda}\\
M_{{\rm g}2}({\rm d}t\,{\rm d}t')&=&\left[n^2 +n\delta(t-t')\right]{\rm d}t\,{\rm d}t',\label{PoissM}
\end{eqnarray}
where $n$ is the mean rate of events. Other parameters are treated as independent marks with common
distribution $G({\rm d}\phi\,{\rm d}r\,{\rm d}\mbox{\boldmath $\xi$})$. The second-order measure
of the process has a form
\begin{eqnarray}
M_2({\rm d}t\,{\rm d}\phi\,{\rm d}r\,{\rm d}\mbox{\boldmath $\xi$}
\,{\rm d}t'\,{\rm d}\phi'\,{\rm d}\mbox{\boldmath $\xi$}')
&=&\left[n^2G({\rm d}\phi\,{\rm d}r\,{\rm d}\mbox{\boldmath $\xi$})G({\rm d}\phi'\,{\rm d}r'\,{\rm d}\mbox{\boldmath $\xi$}')
+nG({\rm d}\phi\,{\rm d}r\,{\rm d}\mbox{\boldmath $\xi$}) \right.\nonumber\\
&\times&\left.\delta(t-t')\delta(\phi-\phi')\delta(r-r')
\delta(\mbox{\boldmath $\xi$}-\mbox{\boldmath $\xi$}')\right]{\rm d}t{\rm d}t'.
\label{somd}
\end{eqnarray}
For the power spectrum we obtain this general formula,
\begin{eqnarray}
S(\omega)&=&4\pi^2n\sum\limits_{k=-\infty}^\infty\sum\limits_{l=-\infty}^\infty 
\int\limits_{\mathcal{K}}c_k(r) c^*_l(r)e^{i(l-k)\phi}
\mathcal{F}\left[I(t-\delta t(r,\,\phi),\,\mbox{\boldmath $\xi$})\right]\left(\omega-k\Omega(r)\right)
\nonumber\\ 
&\times&
\mathcal{F}^*\left[I(t-\delta t(r,\,\phi),\,\mbox{\boldmath $\xi$})\right]\left(\omega-l\Omega(r)\right)
G({\rm d}\phi\,{\rm d}r\,{\rm d}\mbox{\boldmath $\xi$}).
\label{PoissPSD}
\end{eqnarray}

\subsection{Introducing a relationship among spots (Hawkes process)}
The assumption that the spots are mutually statistically independent
seems to be a reasonable first approximation, however, the actual
ignition times and spot parameters should probably depend on the history
of a real system. As an example of such non-Poissonian process, we
calculated the power-spectral density (PSD) for a model in which the
spot ignition times are distributed according to the Hawkes \citep{Hawkes:1971}
process.

The Hawkes process consists of two types of events. Firstly, new events are
generated by Poisson process operating with the intensity $\lambda$. 
Secondly, an existing
event with ignition time $t_a$ can give birth to new event at time $t$
according to Poisson process with varying intensity $\mu(t-t_a)$. So the mean number of events found at
time $t$ is
\begin{equation}
m(t)=\lambda+\sum\limits_{i,\,t_i< t}\mu(t_i)=\lambda+\int\mu(t)N({\rm d}t).
\end{equation}
For a stationary process the first moment density is constant. Averaging
both sides of the previous equation we find,
\begin{equation}
m_1=\frac{\lambda}{1-\nu},\quad \nu=\int\limits_{-\infty}^\infty\mu(t){\rm d}t.
\end{equation}
Stationarity of the process implies, that the second-order measure
density  can depend only on the difference of its arguments. It can be
proven \citep{Daley:2003} that
\begin{equation}
m_{{\rm g}2}(t,t')=c(t-t')+m_{{\rm g}1}^2 +m_{{\rm g}1}\delta(t-t'),
\end{equation}
where the $c(t)$ is an even function.  Thus, for the corresponding marked 
process with  independent marks we find $M_2({\rm d}t\,{\rm d}\phi\,{\rm d}r\,{\rm d}\mbox{\boldmath $\xi$}
\,{\rm d}t'\,{\rm d}\phi'\,{\rm d}\mbox{\boldmath $\xi$}')$:
\begin{eqnarray}
M_2&=&\left[\left(\frac{\lambda^2}{(1-\nu)^2}+c(t-t')\right)G({\rm d}\phi\,{\rm d}r\,{\rm d}\mbox{\boldmath $\xi$})G({\rm d}\phi'\,{\rm d}r'\,{\rm d}\mbox{\boldmath $\xi$}')
\right. \nonumber\\
&&\left.+\frac{\lambda}{1-\nu}G({\rm d}\phi\,{\rm d}r\,{\rm d}\mbox{\boldmath $\xi$})\delta(t-t')\delta(\phi-\phi')\delta(r-r')
\delta(\mbox{\boldmath $\xi$}-\mbox{\boldmath $\xi$}')\right]{\rm d}t{\rm d}t'.
\label{somdHawk}
\end{eqnarray}
This second-order measure is almost identical to that of the Poissonian process
(there is only one additional term associated with the function $c(t)$).
The resulting PSD is
\begin{eqnarray}
S(\omega)&=&4\pi^2\frac{\lambda}{1-\nu}
\sum\limits_{k=-\infty}^\infty\sum\limits_{l=-\infty}^\infty 
\int\limits_{\mathcal{K}}e^{i(l-k)\phi}c_k(r) c^*_l(r)
\mathcal{F}\left[I(t-\delta t(r,\,\phi),\,\mbox{\boldmath $\xi$})\right]\left(\omega-k\Omega(r)\right)
\nonumber\\
&\times&\mathcal{F}^*\left[I(t-\delta t(r,\,\phi),\,\mbox{\boldmath $\xi$})\right]\left(\omega-l\Omega(r)\right)
G({\rm d}\phi\,{\rm d}r\,{\rm d}\mbox{\boldmath $\xi$})\;+\;4\pi^3\mathcal{F}\left[c(t)\right](\omega)\nonumber\\
&\times&\sum\limits_{k=-\infty}^\infty
c_k(r) \int\limits_{\mathcal{K}}e^{-ik\phi}
\mathcal{F}\left[I(t-\delta t(r,\,\phi),\,\mbox{\boldmath $\xi$}')\right]\left(\omega-k\Omega(r)\right)
G({\rm d}\phi\,{\rm d}r\,{\rm d}\mbox{\boldmath $\xi$})
\nonumber\\
&\times&	\sum\limits_{l=-\infty}^\infty \int\limits_{\mathcal{K}'}e^{il\phi'}
c_l^*(r') \mathcal{F}^*\left[I(t-\delta t(r',\,\phi'),\,\mbox{\boldmath $\xi$})\right]
\left(\omega-l\Omega(r')\right)G({\rm d}\phi'\,{\rm d}r'\,{\rm d}\mbox{\boldmath $\xi$}').
\label{HawkPSD}
\end{eqnarray}

The function $c(t)$  can be calculated from the mean number of secondary events $\mu(t)$. 
Assuming $\mu(t)=\nu\alpha\exp(-\alpha t)\theta(t)$ we obtain
\begin{equation}
c(t)=\frac{\lambda\alpha\nu(1-\nu/2)}{(1-\nu)^2}\exp(-\alpha(1-\nu)|t|).
\end{equation}

\begin{figure*}
\includegraphics[width=0.49\textwidth]{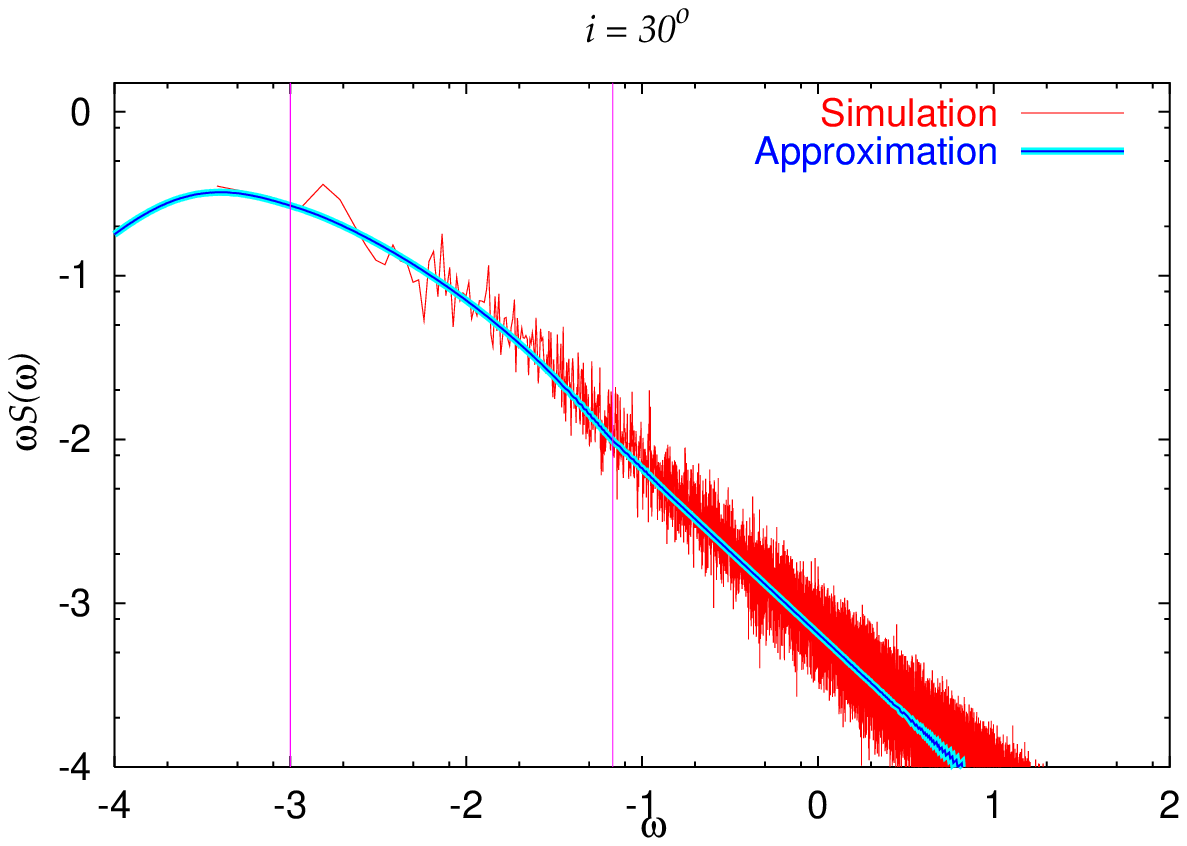}
\includegraphics[width=0.49\textwidth]{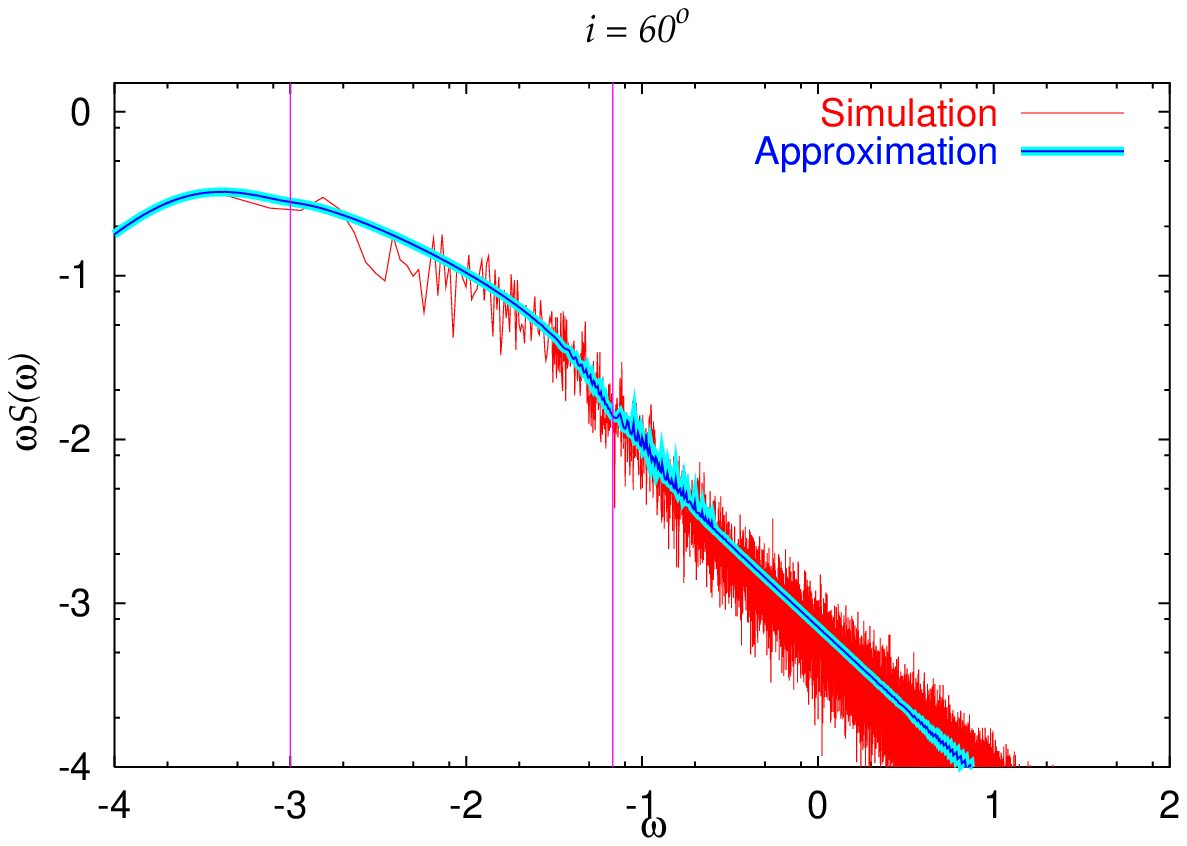}\\
\includegraphics[width=0.49\textwidth]{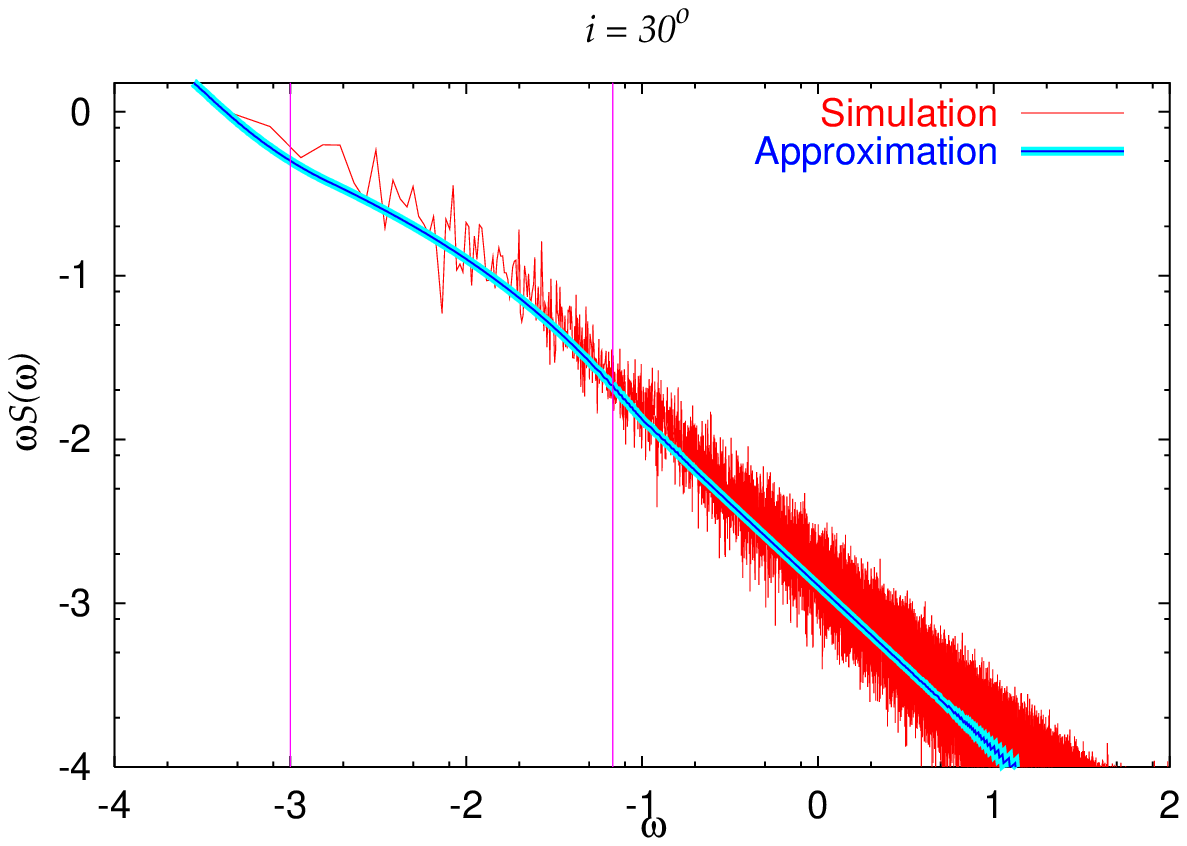}
\includegraphics[width=0.49\textwidth]{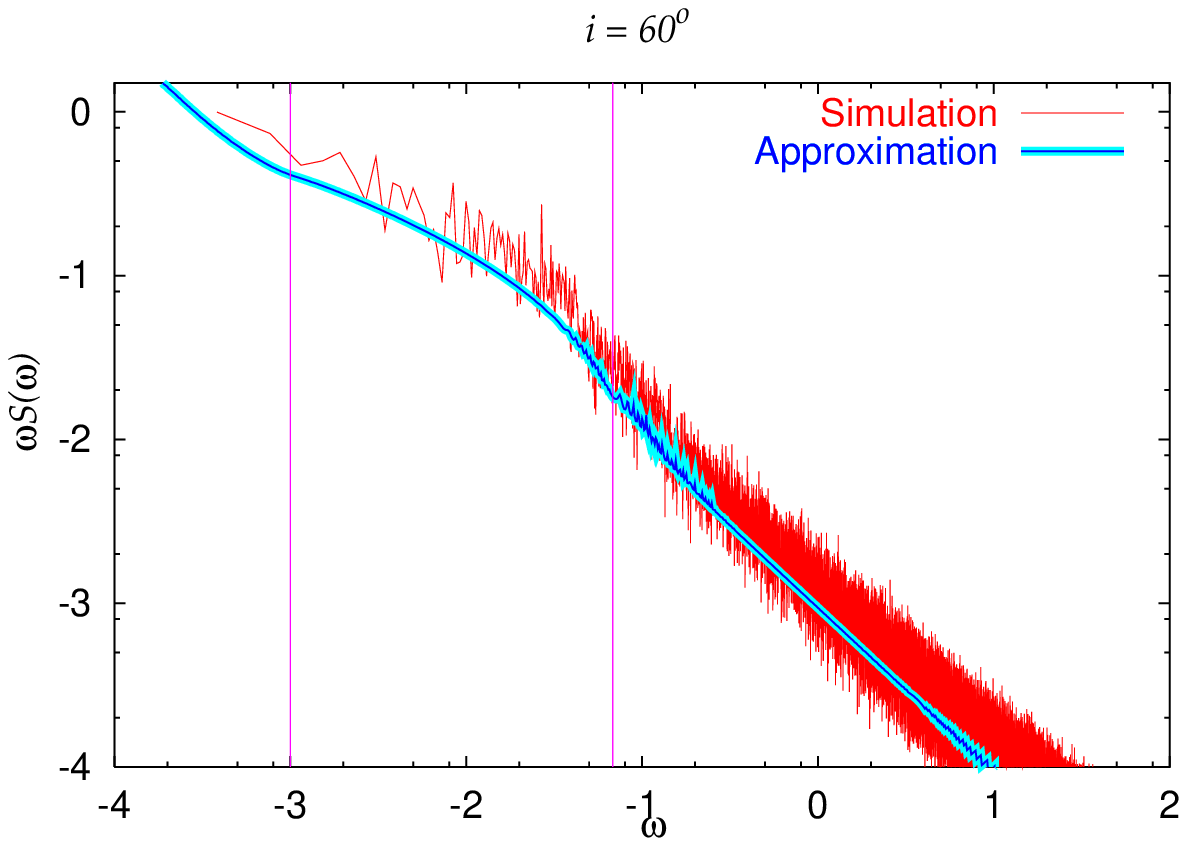}
\caption{
Power spectra from the spot model driven by the Poisson process (top
row) and the Hawkes process (bottom row), calculated for spots orbiting
and evolving on the surface of a thin accretion disc 
($r_{\rm{}in}=6$, $r_{\rm{}out}=100$ gravitational radii). Two values
of observer's inclination $\theta_{\rm o}$ are shown for comparison. The red 
(thin, noisy) curve is a result of direct numerical simulation. Blue 
(thick, continuous) curves are the analytical approximations based on
eqs.~(\ref{PoissPSD}) and (\ref{HawkPSD}), respectively. We assumed
probability density function $\rho(\tau) \propto 1/\tau$. The magenta
(vertical) lines denote the Keplerian orbital frequency $\Omega(r)$ at
the inner and the outer edges of the disc. One can see that the Hawkes'
process tends to enhance the  low-frequency part of PSD and shift the
break frequency towards lower values, below $\Omega(r_{\rm{}out})$.}
\label{psds}
\end{figure*}

It is interesting to notice that the above-given formal approach 
can actually provide a useful analytical formula to approximate
the power spectrum. Figure~\ref{psds} shows exemplary PSD which
were obtained by (i)~direct
computations of the light-curve and the resulting PSD, and by 
(ii)~the semi-analytical approach with Poissonian and Hawkes
processes.

\section{Conclusions}
We have studied the properties of power spectral density within the
model of accretion disc variability driven by orbiting spots. The origin
and evolution of spots were described in terms of Poissonian and
Hawkes' processes. The latter belongs to a category of avalanche models.
We developed an analytical approximation of PSD and
compared it with our numerical results from light-curve simulations. In
this way we were able to demonstrate the precision of formulae
(\ref{PoissPSD}) and (\ref{HawkPSD}). The analytical approximation
evaluates very fast and provides the main trend of the PSD shape while
avoiding the noisy form of the numerically simulated spectra. Our
approach allows us to investigate the resulting PSD as a function
of the assumed type of process, which describes creation of parent spots
and the subsequent cascades of daughter spots. In particular, we can
investigate the predicted PSD slope at different frequency ranges and we
can locate the break frequency depending on the model parameters.
 
The resulting PSD can be approximated by a broken power-law. For every
stationary process the quantities $S(0)$ and $\int_0^\infty
S(\omega){\rm d}\omega$ are finite. Therefore, the function $S(\omega)$
flattens ($S(\omega)\approx \omega^0$) near $\omega=0$ and it
must decrease faster than $1/\omega$ at high frequencies. Power-spectra
generated by the spot model behave in this way. The low-frequency limit
is a constant, whereas the-high frequency behaviour depends mainly on
the shape of the spot emission profile, $I(t,\,\mbox{\boldmath
$\xi$}_j)$. In our calculations the emissivity was a decaying
exponential and the slope was equal to $-2$ at high frequencies. The
most interesting part of the spectra in between those two limits is
influenced by both the emissivity profile and the underlying process. 

\ack
We thank the Workshop participants for helpful
comments and the Academy of Sciences for financial support (grant
GAAV 300030510).

\end{document}